\documentclass[preprintnumbers,amsmath,amssymb]{elsart}

\usepackage{graphicx}
\usepackage{dcolumn}
\usepackage{bm}

\usepackage{float}
\usepackage{afterpage}

\begin{document}

\begin{frontmatter}

\title{Auger Electron Cascades in Water and Ice}
\author{Nicu\c{s}or T\^\i mneanu},
\author{Carl Caleman},
\author{Janos Hajdu},
\author{David van der Spoel\thanksref{label2}}
\address{Department of Cell and Molecular Biology, Biomedical Centre, 
Box 596, Uppsala University, SE-75124 Uppsala, Sweden}
\thanks[label2]{email: spoel@xray.bmc.uu.se}

\date{\today}

\begin{abstract}
Secondary electron cascades can induce significant ionisation in
condensed matter due to electron-atom collisions. This is of
interest in the context of diffraction and imaging using X-rays, where
radiation damage is the main limiting factor for achieving high
resolution data. Here we present new results on electron-induced
damage on liquid water and ice, from the simulation of Auger electron
cascades. We have compared our theoretical estimations to the
available experimental data on elastic and inelastic electron-molecule
interactions for water and found the theoretical results for elastic
cross sections to be in very good agreement with experiment. As a
result of the cascade we find that the average number of secondary
electrons after 100 fs in ice is about 25, slightly higher than in water,
where it is about 20. The difference in damage between ice and water is
discussed in the context of sample handling for biomolecular systems.
\end{abstract}

\end{frontmatter}

\newpage
\section{Introduction}
High energy X-rays damage matter, mainly through the photoelectric
effect.  A photon is absorbed by an atom and a photoelectron is
emitted from the inner (K) shell of the atom.  
In biological molecules, the remaining hollow
ion might relax through Auger decay, in which an electron from a higher
level falls down and a further electron is emitted with an energy
 between 250 eV and 2 keV~\cite{Thompson2001}.
 The basic atomic physics for these processes is well understood
and cross sections for photoionization have been
tabulated~\cite{Veigele73,Hubbell75}. Furthermore, Auger line widths
have been measured giving K-hole life times,
of \textit{e.g.} 11.1 fs (C), 9.3 fs (N), 6.6 fs (O) and 1.3 fs (S) 
~\cite{Krause79}.
 During the photoionization process, the 
photoelectron
may interact with electrons in the valence shell. For elements of 
biological
significance this may lead to the so called shake-up and 
shake-off
effects~\cite{Siegbahn69} in which a further low energy electron  
(10-100 eV)
is emitted. Semi-empirical quantum calculations were used to estimate 
that
such shake-off ionizations occur in 10 - 30\% 
events~\cite{Persson2001a}.

The advent of X-ray free electron lasers (FEL) will enable
a whole range of new experiments in physics, chemistry and
biology 
within a few years~\cite{Wiik97,Hajdu2000,Winick95}. 
In the category of biological applications, we envision
performing structural studies on large biomolecules, biomolecular
aggregates or nanocrystals. Molecular dynamics simulations of a
protein molecule in a FEL beam are encouraging as to the feasibility
of such experiments~\cite{Neutze2000a}.
Preliminary studies with a simple hydrodynamic model have suggested that 
~\cite{London}
a layer of water or ice may delay the eventual Coulomb explosion of a
biomolecule, such as a protein,  
in a FEL beam ~\cite{Neutze2000a}. The reason for this is that 
water molecules  
at the surface supply 
electrons to neutralize the
charging core of the droplet, where the protein is expected to
be situated, while protons, carrying the positive charge,
 are expelled from the water shell.
This may be advantageous in future experiments 
if the sample 
 is held in water droplets.

In the aftermath of a photoionization event in a liquid or solid,
further electrons will be liberated through collisions between the
Auger electron and the atoms, leading to a cascade of ionizations. Recent
studies ~\cite{Ziaja2002a,Ziaja2001a} 
have made a critical assessment of the total amount of
electrons due to an Auger electron of 250 eV 
(carbon)\footnote{Although the actual Auger energy for carbon in diamond is
265 eV~\cite{Yue1996}, we have used the value of 250 eV for comparison with
earlier work.}. 
It was predicted that in diamond,
 between 6 and 13 electrons will be released, while
for amorphous carbon the number is somewhat 
higher due to the 
absence of
a band gap~\cite{Ziaja2002a}.
Here we present results on water, both in liquid and 
solid state.  
The understanding of Auger electron cascades in water is essential 
for modeling the shielding effects water molecules will have
on \textit{e.g.} a protein sample. 
Furthermore, water 
is less dense than diamond 
and therefore more comparable to biomolecules. 
The Auger electron energy for oxygen in a 
water molecule is 507.9 eV ~\cite{Wagner1980}, providing
twice as much energy to the electron than carbon.

\section{Theory}
To calculate the cross sections for scattering of electrons on atoms 
used in the molecular dynamics model,
we have used a solid state approach, assuming a neutral crystal, or, 
in the case of liquid water, a neutral cluster with periodic boundary
conditions. The same approach
has been used earlier by Ziaja \textit{et~al.}
~\cite{Ziaja2002a,Ziaja2001a}. Under this assumption, there are 
three different interactions that a free electron
in the medium can undergo:
(i) elastic collision with atoms,
(ii) inelastic collision with atoms or 
(iii) recombination with atoms or ions.
The probability of electron-ion 
recombination is low on the time scale
considered here (${t \sim 100}$~fs) and therefore neglected 
~\cite{Landau}. Inelastic collision of electrons with ions are not considered,
since the number of ionizations caused by a photoelectron or
single primary Auger electron is much smaller than the total number
of atoms in the sample. 
Scattering of electrons on phonons and the vibrational cross section
are not considered here, though they become relevant for low energies 
($E\leq 10$ eV)~\cite{Michaud2003}.

\subsection*{Elastic scattering}
The cross section for the elastic 
scattering of electrons on atoms is calculated using programs from 
the Barbieri/van Hove Phase Shift package~\cite{Hove}. The
calculations are based on the partial wave expansion 
technique, using the muffin-tin potential approximation.
 The expression 
for the elastic cross section as a function of the electron energy $E$, is   
\begin{equation}
\label{cross_section_elast}
\sigma_{\mathrm{elastic}}(E)=\frac{2\pi\hbar^2}{Em_{\mathrm{e}}}
		\sum_l(2l+1)\sin^2\delta_l,
\end{equation}
where $\delta_l$ is the phase shift of each partial wave $l$.
For further details on the elastic cross section 
calculations see reference~\cite{Ziaja2001a} and 
references therein. 

\begin{figure}[H]
\centering              
\includegraphics[width=12 cm]{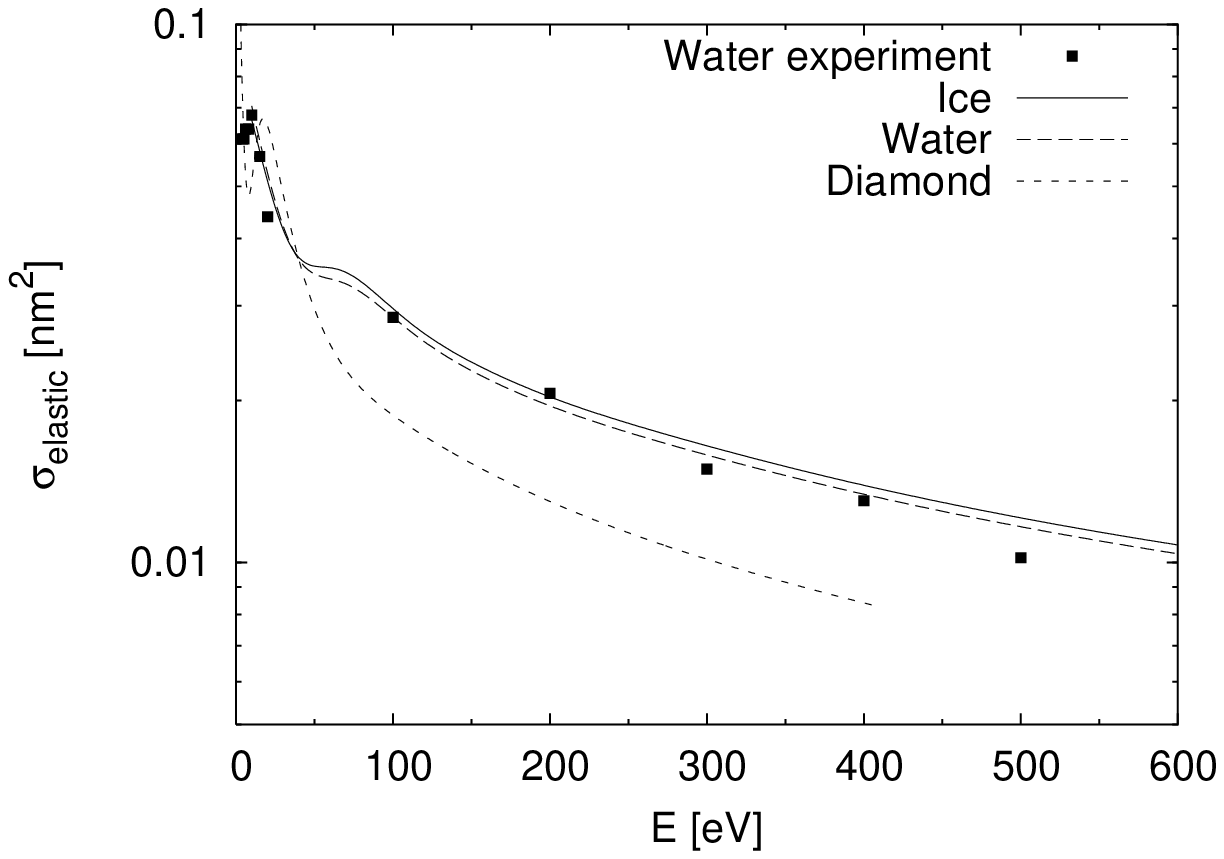}            
\caption{Timneanu et al.} 
\label{figure:elastic_cross_section_ice/water}
\end{figure}

The energy of an elastically scattered electron is conserved, and the
only effect the collision has on the system is that the direction of
the electron velocity changes.  For the calculations of the cross
sections in liquid water we have used a unit cell with hydrogen bonded
water molecules at liquid density (0.9965~$g/cm^3$), while for ice calculations we have
used hexagonal ice, $I_{\mathrm{h}}$ (density 0.9197~$g/cm^3$). 
The results of our calculations
are shown in Fig. $\ref{figure:elastic_cross_section_ice/water}$
together with results for diamond~\cite{Ziaja2001a}.  The results
compare very well with the experimental data for electron scattering
on water molecules.

\nocite{Hove,Ziaja2001a,Danjo,Katase}

\subsection*{Inelastic scattering}
\begin{figure}[H]
\centering              
\includegraphics[width=12 cm]{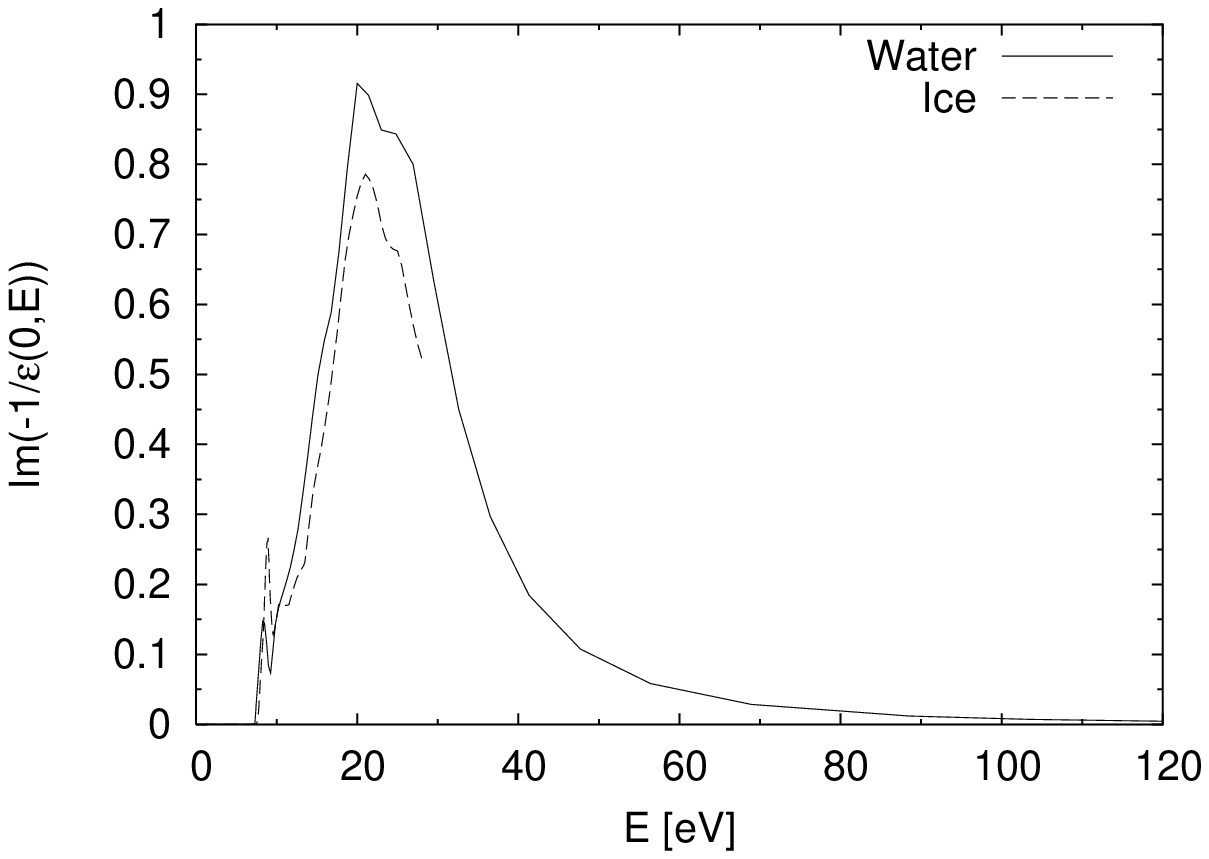}            
\caption{Timneanu et al.} 
\label{figure:dielectric_function_ice/water}
\end{figure}

A correct treatment of the inelastic scattering is more complicated
 and, to our knowledge, there is no complete 
method to describe this. We followed the approach presented in our previous
work~\cite{Ziaja2001a}. 
We have used two different optical data models,
 the Ashley model~\cite{Ashley1990,Ashley1991} and the
Tanuma, Powell and Penn model (TPP2)~\cite{Penn1976,Tanuma1988},
together with the Lindhard dielectric
function $\varepsilon(q,\omega)$~\cite{Fermi1924,Bethe1930,Lindhard1954}
for the calculations of the inelastic cross section
\begin{equation}
\label{cross_section_inelast}
\sigma_{\mathrm{inelastic}}(E)\sim\frac{1}{E}
	\int_{\mathrm{0}}^{\infty}\mathrm{d}\omega
			\int^{q_{+}}_{q_{-}}\frac{\mathrm{d}q}{q}
		\mathrm{Im}\left(\frac{-1}{\varepsilon(q,\omega)}\right),
\end{equation}
were $\hbar q_{\pm}
=\sqrt{2m_{\mathrm{e}}}\left(\sqrt{E}\pm\sqrt{E-\hbar\omega}\right)$.
The above cross section depends on the knowledge of the dielectric
function, which is only known for photons, \textit{i.e.} for momentum
transfer equal to zero. The two different models, Ashley and TPP2,
present different ways to empirically compensate for that. The Ashley
approach includes the exchange between the incoming electron and the
electrons in the crystal, in accordance with the non relativistic M\o
ller cross section~\cite{Kaku}. The TPP2 model is developed for
calculating the differential inelastic mean free path of electrons in
a solid.  A more thorough discussion of the two models can be found in
ref.~\cite{Ziaja2001a}.  From the absorption coefficient for liquid
water ~\cite{Palik}, we deduced the dielectric function
$\varepsilon(0,\omega)$ for zero momentum transfer. For ice we used
experimental data ~\cite{Warren1984} for energies up to 30 eV, and for
higher energies we have extrapolated the values assuming the shape of
the curve to be the same as for liquid water (Fig.
$\ref{figure:dielectric_function_ice/water}$).

\nocite{Ashley1990,Ashley1991,Penn1976,Tanuma1988,Ziaja2001a}

\begin{figure}[H]
\centering              
\includegraphics[width=12 cm]{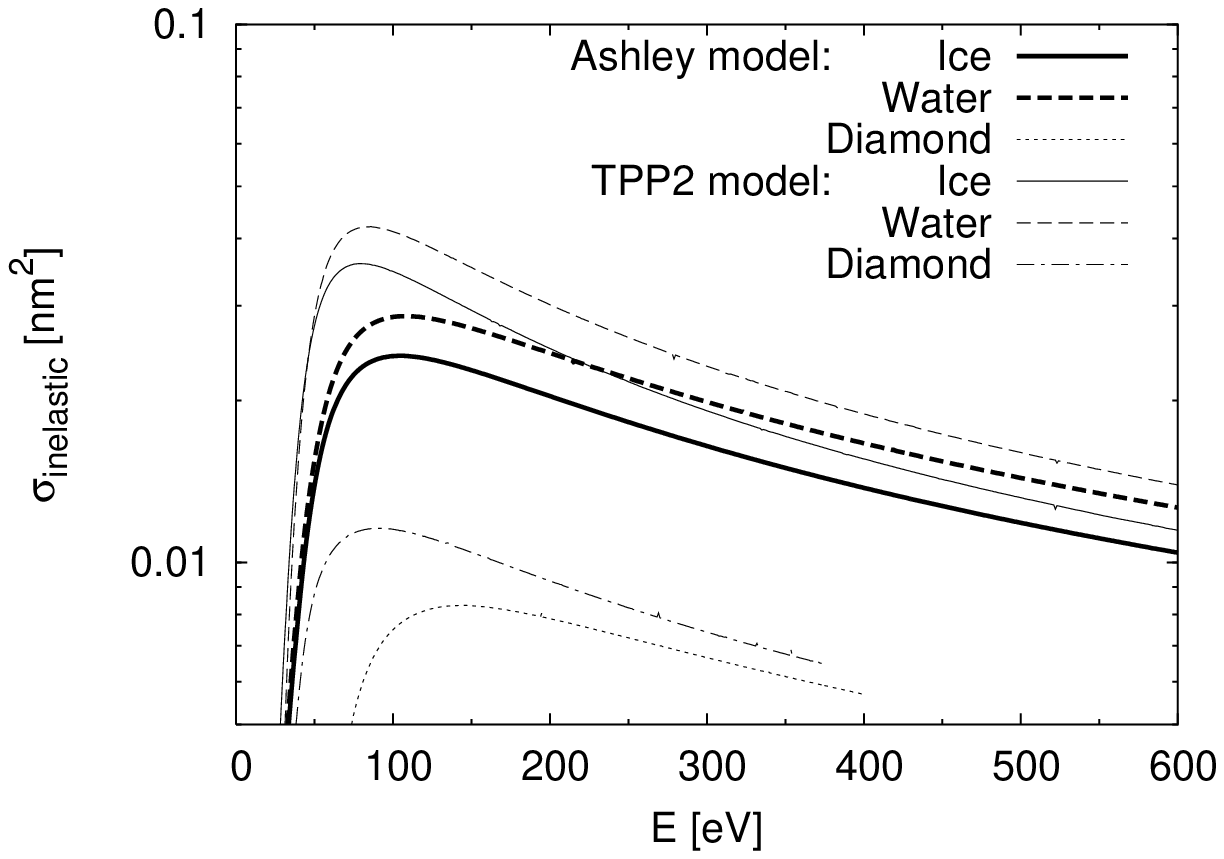}  
\caption{Timneanu et al.}  
\label{figure:inelastic_cross_section_ice/water/diamond}
\end{figure}

An electron that scatters inelastically will change its path
and loose energy to the lattice and, if the lost energy is high enough 
to excite a bound electron from the valence band to the conduction band,
to the excited electron. For simplicity
we assumed that the energy lost to the lattice is
absorbed into the system and in the present study we have not included
any treatment for holes.
The band gaps of water and ice we have used for the calculations were 
$E_{\mathrm{gap}}^{\mathrm{water}}=8.7$ eV ~\cite{Shibaguchi1977} and   
$E_{\mathrm{gap}}^{\mathrm{ice}}=7.8$ eV ~\cite{Petrenko1993}. For water 
the Fermi level $E_{\mathrm{F}}^{\mathrm{water}}=14.85$ eV is taken
from experimental data ~\cite{Shibaguchi1977}, and for ice 
$E_{\mathrm{F}}^{\mathrm{ice}}=11.74$ eV was calculated using the 
free electron approximation with the density taken from reference
 ~\cite{Rottger1994}. The calculations were performed at temperature
$T_{\mathrm{ice}}=270$ K and $T_{\mathrm{water}}=300$ K.
The inelastic cross sections for ice and water are shown in Fig.
 $\ref{figure:inelastic_cross_section_ice/water/diamond}$ together 
with those from diamond, taken from reference ~\cite{Ziaja2001a}.

\nocite{Penn1976,Tanuma1988,Bell1983,BEB}

\begin{figure}[H]
\centering              
\includegraphics[width=12 cm]{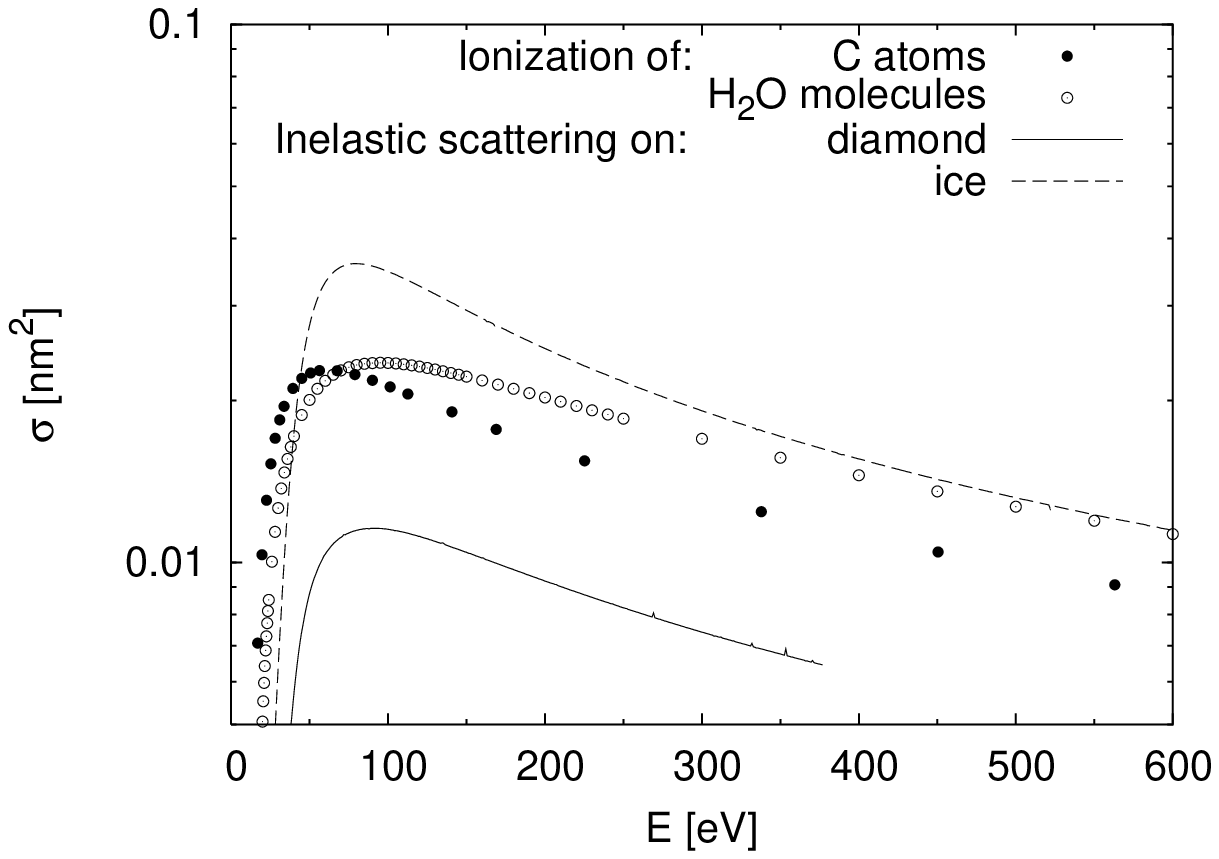}            
\caption{Timneanu et al.} 
\label{figure:campari}
\end{figure}

From the wealth of experimental data for electron impact ionization of
atoms and molecules, models and recommended cross sections have emerged
for different targets. We have compared the theoretical estimates for
inelastic scattering of electrons in solids (diamond, ice), with the recommended
cross sections for carbon~\cite{Bell1983} and a successful model that describes
ionization of water molecules~\cite{BEB} (Fig.  $\ref{figure:campari}$). 
The former represent a fit to the atomic data for several light atoms and 
ions~\cite{Bell1983}, while 
the latter, known as 
the Binary-Encounter-Bethe (BEB) model, 
provides analytical formulae for the ionization cross section
of atomic and molecular orbitals by summing up all contributions. 
The BEB model was found to  reproduce  experimental data for 
small atoms and a variety of molecules very well~\cite{BEB}.

\nocite{Ashley1990,Ashley1991}

We find that, in the case of diamond, the inelastic cross section is lower 
than the ionization of single atoms, reflecting the nonlocalized nature of the
electrons in the solid. This difference could prove significant for the 
estimation of electron interaction in solids, if the solid loses its 
crystalline structure (due to Coulomb explosion while exposed to a FEL
beam). For ice and water, however, we find a smaller difference between
the ionization cross sections~\cite{BEB} and our theoretical estimates
(Fig.  $\ref{figure:campari}$), and with the opposite sign.

\section{Methods}
Classical simulations of the electron trajectories were performed,
where Newton's equations of motion were solved using a Leap-Frog
integration scheme~\cite{Berendsen86b} with a time step of ${\Delta t=1}$ 
attoseconds. The
lattice, be it diamond, water or ice, was modeled implicitly through
the cross sections for collisions.  Elastic as well as inelastic
collisions were modeled as stochastic processes, where the probability
$P(E)$ for an event was determined from:
\begin{equation}
P(E) ~=~ 
 \frac{\rho N_{\mathrm{av}}}{m} \sigma(E) \cdot |v_{\mathrm{e}}(E)| \cdot \Delta
 t ,
\end{equation}
where $\rho$ is the
density of the material, $N_{\mathrm{av}}$ is Avogadro's number, $m$ 
is the mass
of the atom (C in diamond), respectively water molecule, and $v_{\mathrm{e}}$ 
is the velocity of
the electron. ${\Delta t}$ was
chosen such that ${P\ll 1}$ in all cases. 
At each time step in the simulation  
the following processes occur:
\newline
(1) An electron with a certain kinetic energy ${E_{\mathrm{k}}}$ travels 
through the sample. 
The scattering probability (both inelastic and elastic) 
$P(E_{\mathrm{k}})$ is compared
to a random number ${0\le r < 1}$. If no scattering 
occurs, the velocity is kept unchanged.
\newline
(2a) If the electron scatters elastically it changes 
its momentum according to a probability function derived from
the differential cross section (the formulae in ~\cite{Ziaja2001a}
were applied for water and ice). No momentum transfer to the lattice
has been considered at this point.
\newline 
(2b) In the case of inelastic scattering, the electron will
lose energy ${E_{\mathrm{lost}}}$ and change its path. If 
${E_{\mathrm{lost}} > E_{\mathrm{gap}}}$
an additional electron is liberated from the valence band with 
${E_{\mathrm{k}}=E_{\mathrm{lost}}-E_{\mathrm{gap}}}$.
 If ${E_{\mathrm{lost}} < E_{\mathrm{gap}}}$, the 
energy is transfered to the lattice. In both cases the
momentum is changed based on the differential cross section
calculated for water and ice using formulae from~\cite{Ziaja2001a}.
\newline
(3) The electron position is integrated.
\newline
The spatial electron dynamics program is part of the
GROMACS software package~\cite{Lindahl2001a}.

\section{Results and Discussion}
The classical dynamics simulations of the electrons produced the same results for
diamond as our earlier Monte Carlo simulations~\cite{Ziaja2002a},
showing that the implementation of the algorithm is correct, and that
the integration time step was small enough. In principle, the model
could be enhanced by adding electrostatic interactions between
electrons and ions~\cite{Jurek2003}. It is however not trivial to model
the interactions between electrons and ions classically without severe
approximations. Since we have only simulated single Auger electron
trajectories, and the cross sections were determined for neutral
systems, we have deemed it unnecessary to add the Coulomb
term. Finally, we note that correct treatment of electron/ion or
electron/hole systems can be performed using \textit{e.g.} time-dependent density
functional calculations, but it is complicated, and not computationally
feasible for macroscopic systems~\cite{Marques2001a}.

\subsection*{Ionization rates}
\begin{figure}[H]
\centering              
\includegraphics[width=12 cm]{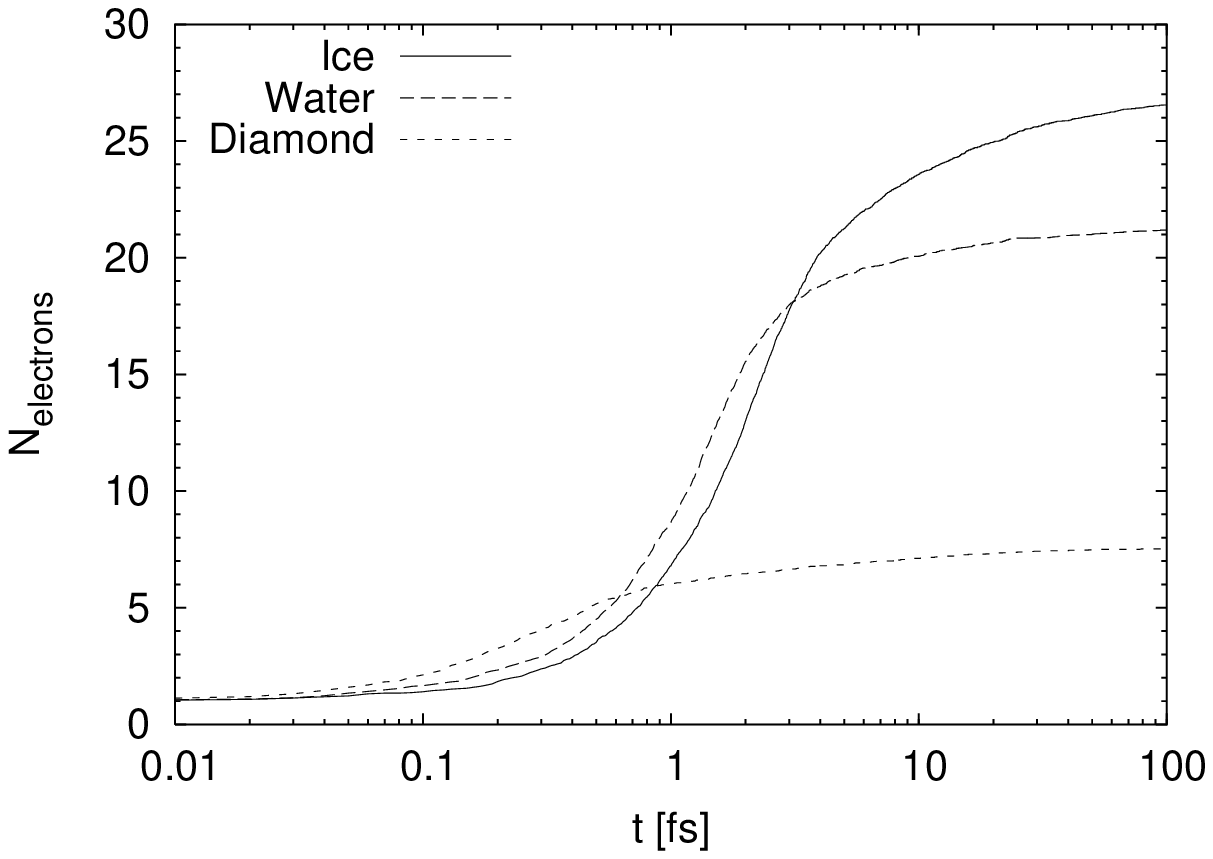}            
\caption{Timneanu et al.} 
\label{figure:nions}
\end{figure}

We find that an Auger electron of 500 eV will produce typically
between 5 and 10 secondary electrons in the first femtosecond in ice and water,
a result which is comparable with the average number of ionization events in
diamond from a 250 eV Auger electron. In ice and water, however, saturation
appears much later, after about 100 fs, when the average number of ionizations
is around 20 for liquid water and 25 for ice.
These results are shown in Fig.~\ref{figure:nions}, which not only compares the 
ionization rates for ice, water and diamond, but also shows that the simulation
of diamond using classical dynamics reproduces previous results obtained with 
a Monte Carlo method well~\cite{Ziaja2002a}.
We note that, although the cross sections for ice and water are rather similar,
the difference in the number of ionizations in the
Auger electron cascade is determined mainly by the
the approximately 1 eV difference in the band gap between liquid water and ice.


\subsection*{Energy distribution}
\begin{figure}[H]
\centering              
\includegraphics[width=12 cm]{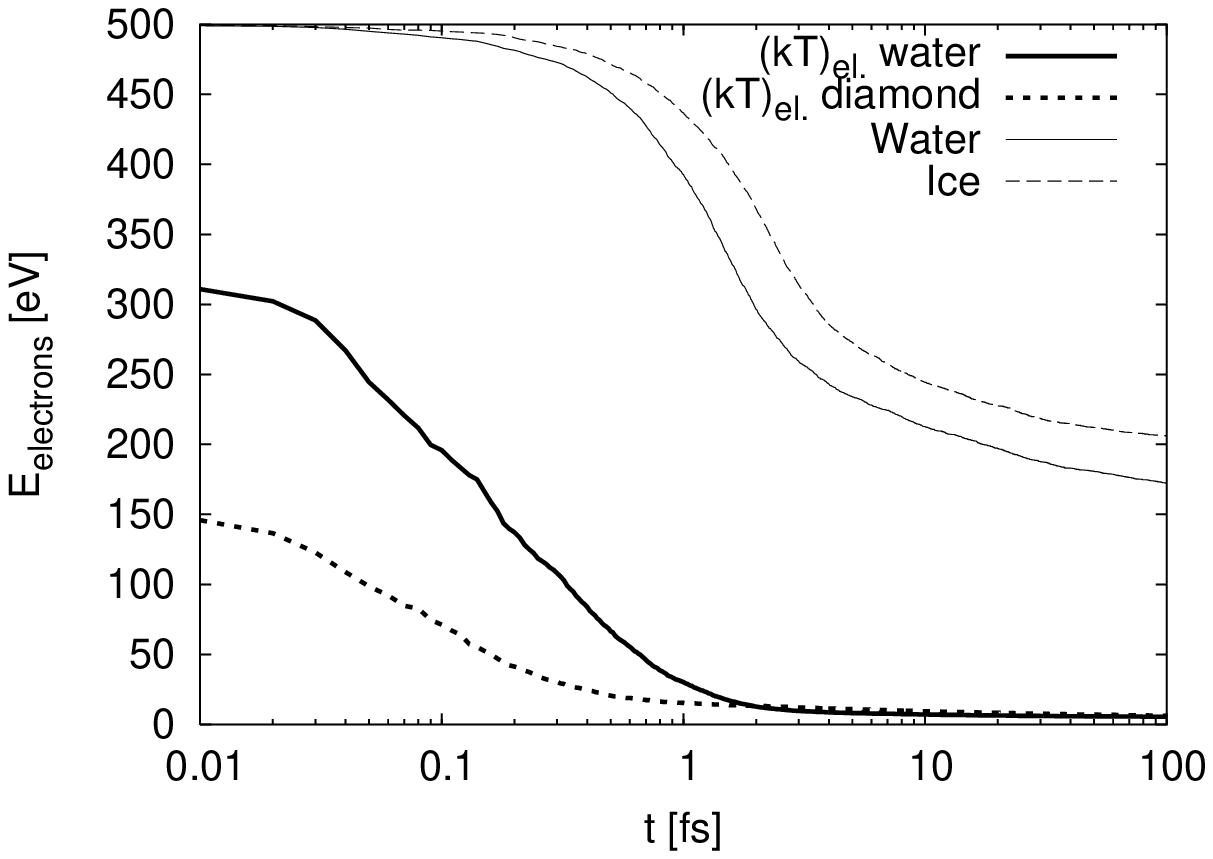}            
\caption{Timneanu et al.} 
\label{figure:energy}
\end{figure}

The total energy of the electrons shows a substantial drop  during the first 10
femtoseconds, with the rest of the energy being transfered to the lattice
(Fig.~\ref{figure:energy}). The electron gas is cooling down very rapidly,
due to the large number of ionizations, and the average temperature of the
electrons $kT_{\mathrm{el}}$ is similar for ice and water. The energy lost to the lattice
is more than 50\% of the initial Auger electron energy.
For comparison, the results obtained
for diamond are also shown, consistent with the previous 
calculations~\cite{Ziaja2002a}.

\begin{figure}[H]
\centering              
\includegraphics[height=8 cm]{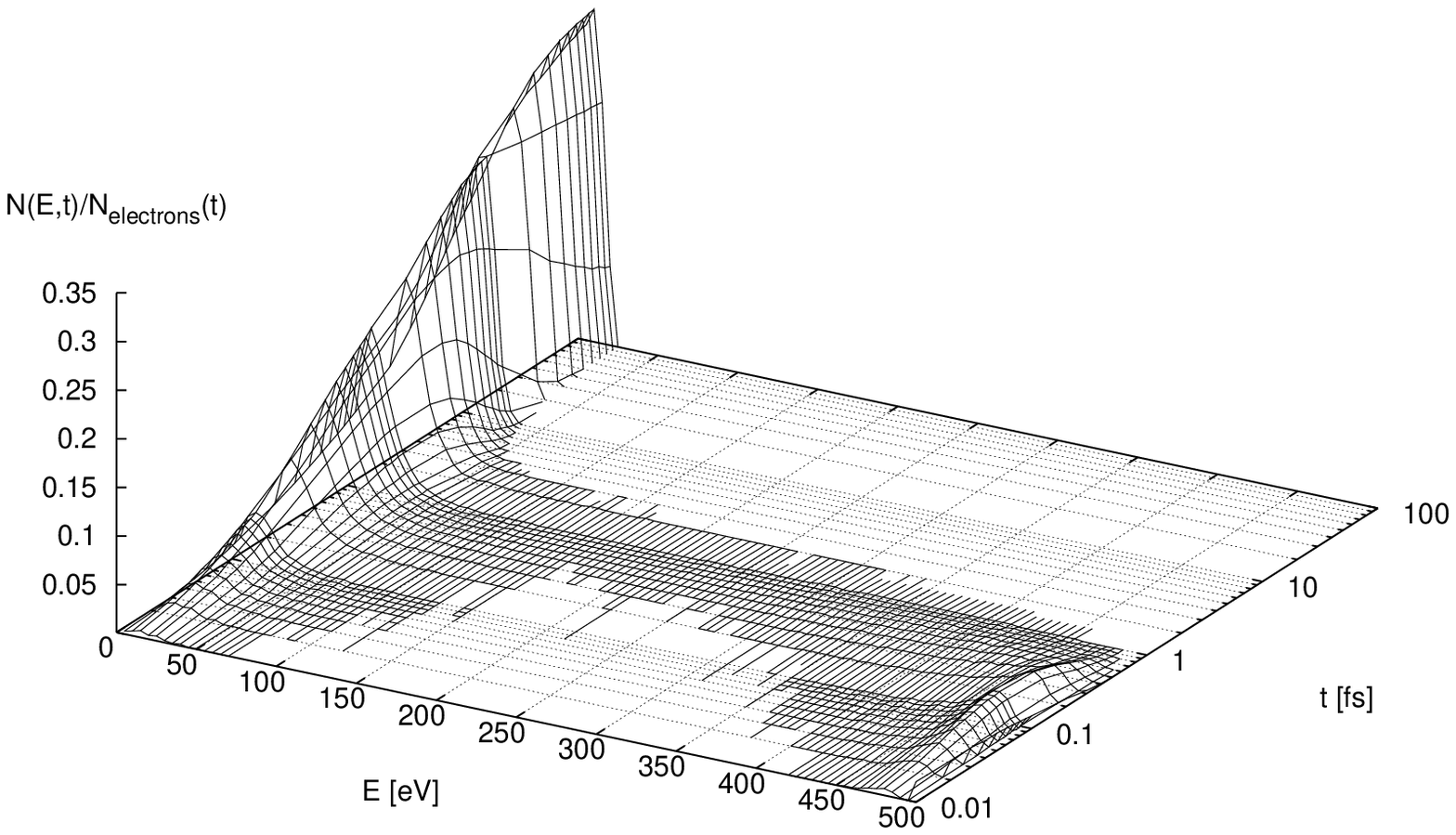}            
\caption{Timneanu et al.} 
\label{figure:3dplot}
\end{figure}

The electron energy distribution as a function of time
in the cascade (Fig.~\ref{figure:3dplot}) is particularly interesting. In the early
stages of the cascade (of the order of 0.1 fs), the electron population is divided
in two sets, initial Auger electrons which are highly energetic 
(450-500 eV) and the first secondary electrons of lower energy. 
At the critical time of about 1 fs, the population  of high energy electrons 
fades away, leaving a spread of electrons with intermediate and low energies.
Subsequent collisions will cool down the electron gas very fast, and after 10 fs
all the electrons will have energies less than 20 eV.  
At such low energies (below 10 eV) and long time scales (100-1000 fs), 
new processes become relevant, such as electron-ion  recombination or 
electron-phonon  scattering, which are not yet included in our model.


\subsection*{Spatial distribution}
\begin{figure}[H]
\centering              
\includegraphics[width=12 cm]{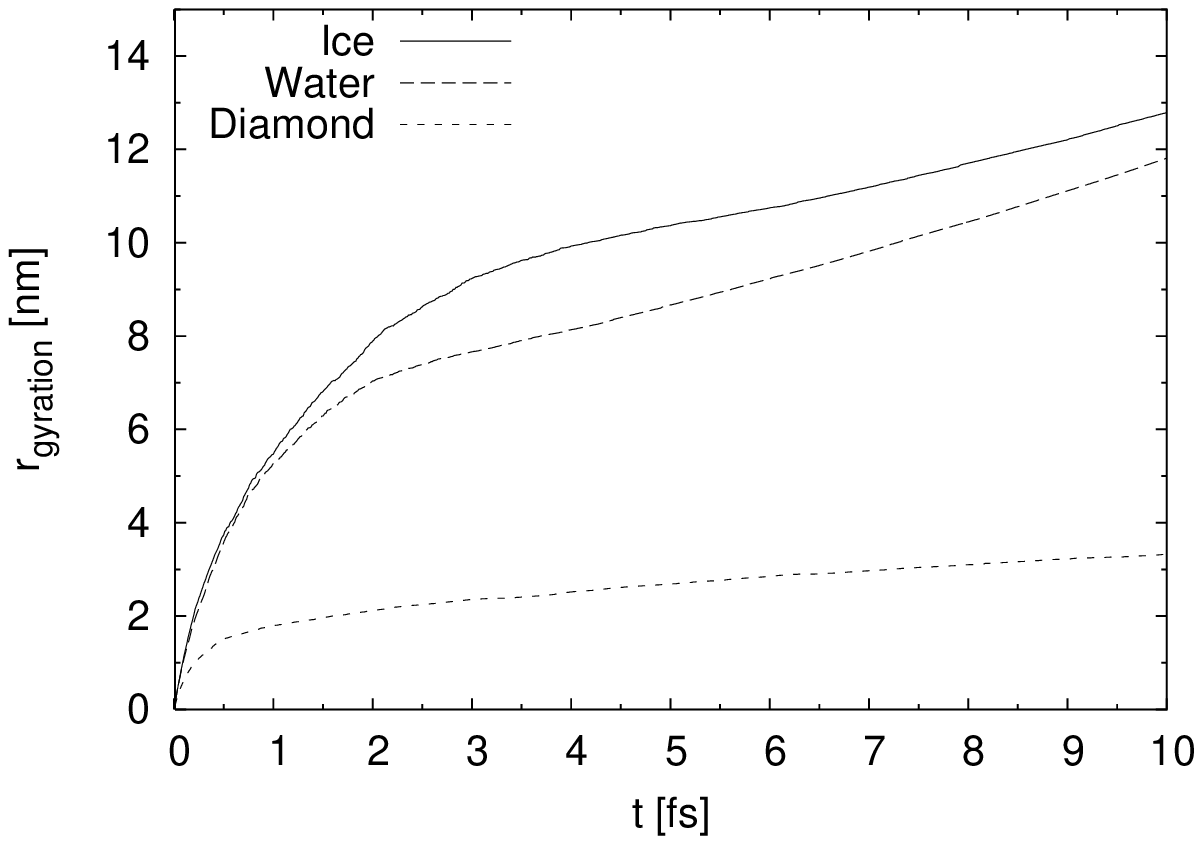}            
\caption{Timneanu et al.} 
\label{figure:gyrate}
\end{figure}

We have performed an analysis of spatial distribution of
electrons in the cascade in our
classical dynamics model for ice, water and diamond.
Fig.~\ref{figure:gyrate}
shows the average gyration radius of the 
electron cloud for ice, water and diamond. The spatial extension of the cloud
in ice and water is three times larger than the one in diamond, due 
to the considerably lower density and the corresponding
longer mean free path an electrons in water/ice.


\section{Conclusion}
An understanding of the behaviour of water and ice exposed to an
intense X-ray beam is important for planned biology experiments with FELs.
Biological macromolecules carry about 30-60\% water with them as structural 
water and water bound as a surface layer.
In the light of results from a hydrodynamic model~\cite{London}, it
was suggested that embedding a sample in water or ice may be
beneficial in the sense that the eventual Coulomb
explosion~\cite{Neutze2000a} could be delayed by the presence of the
water or ice layer. Our results show that in hexagonal ice, $I_{\mathrm{h}}$, 
a slightly higher
number of secondary electrons are liberated as compared to water
(Fig.~\ref{figure:nions}), despite the slightly lower
temperature. Furthermore, the electrons are spread over a larger
volume (Fig.~\ref{figure:gyrate}).  Although the differences are
small, it seems therefore that water is the preferred medium for
encapsulating biomolecules when studied in FEL beams. If the sample is 
embedded in vitrious ice, which could be true in case of a rapid 
freezing, we expect the behaviour to be unaffected compared to water.

\subsection*{Acknowledgments}
We would like to thank
Beata Ziaja, 
Magnus Bergh, 
Jorge Llano, 
Lars Nordstr\"om,  
Richard London, 
Gyula Faigel and
Zolt\'an Jurek
for stimulating discussions.
The Swedish research foundation is acknowledged for financial support.

\bibliographystyle{elsart-num}
\bibliography{../texutil/btmac,../texutil/monster}

\newpage
\section*{Figure captions}
\begin{enumerate}
\item Elastic cross sections for scattering of electrons on ice, water
and diamond. Lines represent the scattering cross section per
scattering center, derived using the Barbieri/van Hove Phase Shift
method~\cite{Hove}, for ice (full line) and water (dashed line).
For comparison, the cross section for diamond taken
from~\cite{Ziaja2001a} is shown (dotted line). The experimental points
represent the elastic scattering of electrons on water molecules as
measured in~\cite{Danjo,Katase}.

\item The imaginary part of the inverse dielectric function,
$\mathrm{Im}(-1/\varepsilon(0,E))$, for water and ice plotted as a
function of the incoming photon energy.

\item Inelastic cross section for electron scattering on ice and water
as a function of the electron energy.  Lines represent the cross
section per scattering center, calculated using two optical data
models, the Ashley model~\cite{Ashley1990,Ashley1991} and the Tanuma,
Powell and Penn model (TPP2)~\cite{Penn1976,Tanuma1988}. For
comparison, the cross sections for diamond taken
from~\cite{Ziaja2001a} are also shown.

\item Comparison between the inelastic cross section for electron
 scattering on solids (ice and diamond) and the ionization cross
 section for electron scattering on single atoms and molecules. Lines
 represent the calculated cross section using TPP2
 model~\cite{Penn1976,Tanuma1988} for diamond (solid line) and ice
 (dashed line).  Points represent the recommended cross section for
 carbon atoms~\cite{Bell1983} (solid points), and the calculated cross
 section Binary-Encounter-Bethe (BEB) model~\cite{BEB} for ionization
 of water molecules (circles).

\item Average number of secondary electrons as a
 function of time, from an initial Auger electron (at time t=0 fs) in
 ice (solid line), water (dashed line) and diamond (dotted
 line). Lines correspond to calculations using Ashley's
 model~\cite{Ashley1990,Ashley1991}.

\item Average energy of the free electrons in an electron cascade coming from an
 initial Auger electron in ice and water as a function of time (thin
 lines).  The rest of the Auger electron energy is transfered to the
 lattice.  Average temperature $kT_{\mathrm{el}}$ of the electron gas
 versus time (thick lines) for water and diamond.

\item Histogram of the energy distribution $N(E,t)/N(t)$ among
electrons at different time steps in ice. At a fixed time, the
histogram is normalized to the number of electrons in the cascade
$N(t)$. Results are shown for an average cascade produced by an Auger
electron in ice.

\item The gyration radius (average distance of electrons from the
cloud center of mass) of the electron cascade as a function of time,
for ice, water and diamond.

\end{enumerate}

\newpage

\newpage

\newpage

\newpage

\newpage

\newpage

\newpage 

\newpage 

\end{document}